\documentclass{appolb}
\usepackage{graphicx}

\newcommand {\snn} {\sqrt{s_{_{\rm NN}}}}
\newcommand {\cme} {{\sc cme}}
\newcommand {\ese} {{\sc ese}}
\newcommand {\poi} {{\sc poi}}
\newcommand {\OS} {{\sc os}}
\newcommand {\Ss} {{\sc ss}}
\newcommand {\RP} {{\sc rp}}
\newcommand {\EP} {{\sc ep}}
\newcommand {\SP} {{\sc sp}}
\newcommand {\PP} {{\sc pp}}
\newcommand {\gos} {\gamma_{_{\rm OS}}}
\newcommand {\gss} {\gamma_{_{\rm SS}}}
\newcommand {\dg} {\Delta\gamma}
\newcommand {\phia} {\phi_{\alpha}}
\newcommand {\phib} {\phi_{\beta}}
\newcommand {\phic} {\phi_{c}}
\newcommand {\psirp} {\Psi_{_{\rm RP}}}
\newcommand {\psisp} {\Psi_{_{\rm SP}}}
\newcommand {\psipp} {\Psi_{_{\rm PP}}}
\newcommand {\bkg} {{\rm Bkg}}
\newcommand {\ru} {{\rm Ru}}
\newcommand {\zr} {{\rm Zr}}
\newcommand {\twp} {{\rm 2p}}
\newcommand {\thp} {{\rm 3p}}
\newcommand {\nf} {{\rm nf}}
\newcommand {\enf} {\epsilon_{\nf}}
\newcommand {\etw} {\epsilon_{2}}
\newcommand {\eth} {\epsilon_{3}}
\newcommand {\fcme} {f_{_{\rm CME}}}
\newcommand {\mean}[1] {\langle#1\rangle}
\newcommand {\vtw}[1] {v_2\{{\rm #1}\}}
\newcommand {\dgbc}[2] {\dg_{\rm #1}\{{\rm #2}\}}

\begin{document}
\title{CME -- Experimental Results and Interpretation
  \thanks{Presented at Quark Matter 2022.}%
}
\author{Fuqiang Wang
  \address{Purdue University, West Lafayette, Indiana 47907, USA\\
    Huzhou University, Huzhou, Zhejiang 313000, China}
}
\maketitle
\begin{abstract}
The experimental status is reviewed on the search for the chiral magnetic effect (\cme) in relativistic heavy-ion collisions. Emphasis is put on background contributions to the \cme-sensitive charge correlation measurements and their effects on data interpretation.
\end{abstract}
  
\section{Introduction}
The vacuum in quantum chromodynamics ({\sc qcd}) possesses a property of the gluon field characterized by the Chern-Simons winding number or the topological charge ($\Delta Q$). Interactions with $\Delta Q\neq0$ gluon field would cause an imbalance in the (anti-)quark chirality. Such an imbalance can lead to charge separation along the direction of a strong magnetic field, a phenomenon called the chiral magnetic effect (\cme)~\cite{Kharzeev:1998kz,Kharzeev:2007jp}.
It is theorized that the \cme\ can arise in non-central heavy-ion collisions, where vacuum fluctuations to $\Delta Q\neq0$ states are possible and where a strong transient magnetic field is present~\cite{Kharzeev:2015znc}, while quantitative predictions are difficult~\cite{Kharzeev:2004ey,Muller:2018ibh}.
Since $\Delta Q\neq0$ explicitly breaks the $\mathcal{CP}$ symmetry, an observation of the \cme\ would not only verify a fundamental property of {\sc qcd} but may also provide a natural solution to the matter-antimatter asymmetry of our universe. 


The magnetic field created in heavy-ion collisions is on average perpendicular to the reaction plane (\RP)~\cite{Skokov:2009qp}. A distinct signature of the \cme\ is back-to-back emissions of opposite-sign (\OS) charged hadrons and collimated emissions of same-sign (\Ss) ones. This motivates the widely used observable, $\dg\equiv\gos-\gss$, the difference in $\gamma_{\alpha\beta}=\mean{\cos(\phia+\phib-2\psirp)}$ between \OS\ and \Ss\ pairs (where $\phia$ and $\phib$ are their azimuthal angles and $\psirp$ is that of the \RP)~\cite{Voloshin:2004vk}. Several other observables~\cite{Ajitanand:2010rc,Tang:2019pbl}
have been proposed; since they are connected to $\dg$~\cite{Choudhury:2021jwd}, only $\dg$ will be reviewed here. 

\begin{figure}[htb]
\centering
\includegraphics[height=3.8cm]{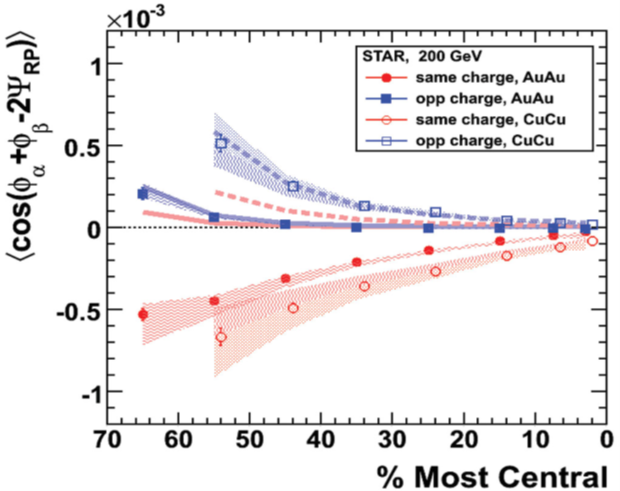}\hspace{1cm}
\includegraphics[height=3.8cm]{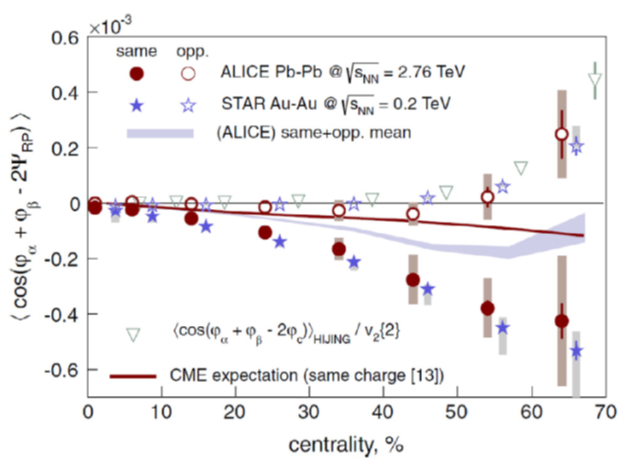}
\caption{First measurements of $\gos$ and $\gss$ in Au+Au and Cu+Cu collisions by RHIC/STAR~\cite{Abelev:2009ac,Abelev:2009ad} (left) and in Pb+Pb collisions by LHC/ALICE~\cite{Abelev:2012pa} (right).}
\label{fig:early}
\end{figure}

{\em Early measurements.}
Figure~\ref{fig:early} shows the first measurements of $\gos$ and $\gss$ at RHIC by STAR~\cite{Abelev:2009ac,Abelev:2009ad,Adamczyk:2013hsi}
and at the LHC by ALICE~\cite{Abelev:2012pa}. Large $\dg$ signals were observed, qualitatively consistent with expectations from the \cme.
Although charge-independent backgrounds have canceled in $\dg$, charge-dependent backgrounds remain~\cite{Voloshin:2004vk,Wang:2009kd,Bzdak:2009fc,Schlichting:2010qia}. 
The larger $\dg$ in Cu+Cu than Au+Au collisions is consistent with such backgrounds which are typically inversely proportional to multiplicity ($N$).
It was warned in the first publications~\cite{Abelev:2009ac,Abelev:2009ad} that {\it ``[i]mproved theoretical calculations of the expectesd signal and potential physics backgrounds ... are essential to understand whether or not the observed signal is due to [\cme]''}

{\em Backgrounds.}
The $\dg$ is ambiguous between a back-to-back \OS\ pair perpendicular to the \RP\ (\cme\ signal) and a collimated one parallel to it (background). Because of elliptic flow ($v_2$), there are more resonances (or generally clusters) thus more \OS\ pairs along the \RP, leading to a background. It arises from the coupling of $v_2$ and genuine two-particle (2p) correlations (part of nonflow)~\cite{Voloshin:2004vk,Zhao:2019hta}:
\begin{equation}
  \dg_{\bkg}=N_{\rm cluster}/(N_{\alpha}N_{\beta})\cdot\mean{\cos(\phi_{\alpha}+\phi_{\beta}-2\phi_{\rm cluster})}\cdot v_{2,{\rm cluster}}\,.
  \label{eq:bkgd}
\end{equation}
Order of magnitude estimate suggests a background level of $0.2/100\times0.5\times 0.1 \sim 10^{-4}$, comparable to the measured $\dg$. 
In fact, thermal and Blast-wave model parameterizations of particle yields and spectra data can reproduce the majority, if not the full, strength of the measurement~\cite{Schlichting:2010qia,ALICE:2020siw}.

\begin{figure}[htb]
  \begin{minipage}{0.49\textwidth}
    \centering
    \includegraphics[width=0.85\textwidth]{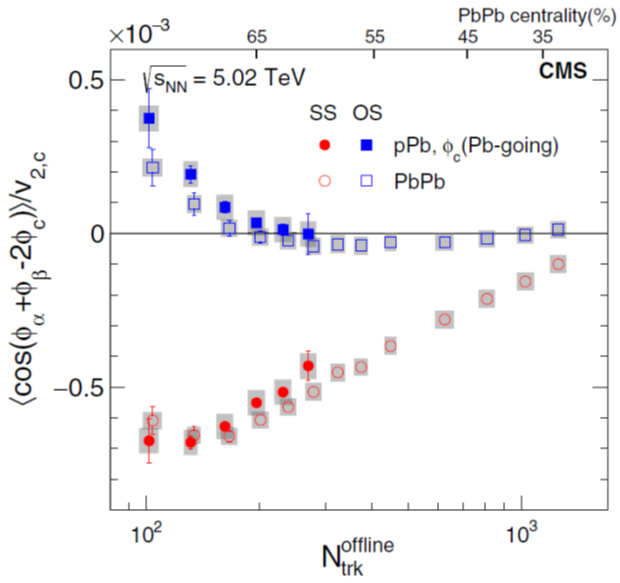}
    \caption{The $\gos$ and $\gss$ in p+Pb and Pb+Pb collisions measured by CMS~\cite{Khachatryan:2016got}.}
    \label{fig:cms}
  \end{minipage}\hfill
  \begin{minipage}{0.49\textwidth}
    \centering
    \includegraphics[width=0.78\textwidth]{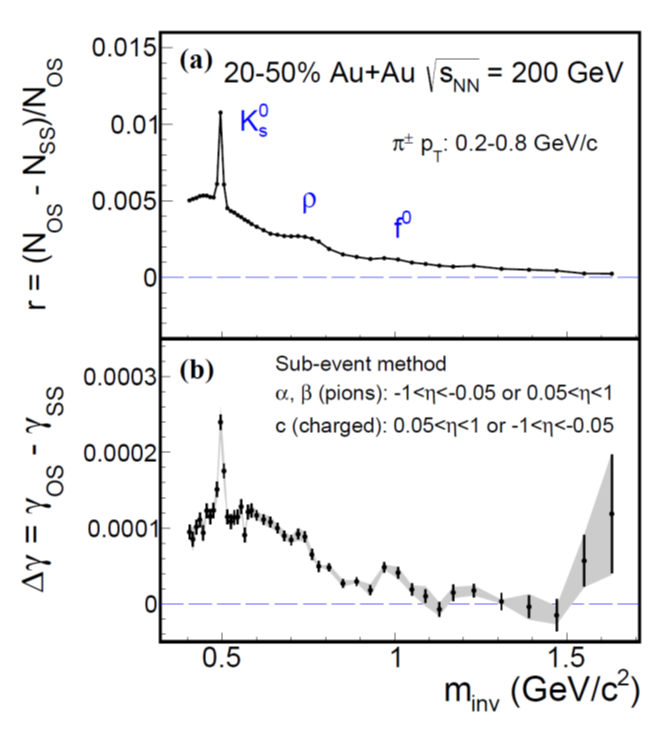}
    \vspace{-3mm}
    \caption{The pair excess $r$ and $\dg$ vs.~$m_{\rm inv}$ in Au+Au collisions by STAR~\cite{Adam:2020zsu}.}
    \label{fig:minv}
  \end{minipage}
\end{figure}

The first experimental indication that background is large is the CMS measurement~\cite{Khachatryan:2016got} in p+Pb collisions being comparable to that in Pb+Pb, as shown in Fig.~\ref{fig:cms}. Similar observation is made in p/d+Au collisions at RHIC~\cite{STAR:2019xzd}. In those small-system collisions, the reconstructed event plane (\EP) is not correlated to the magnetic field direction, hence any \cme\ would not be observable~\cite{Khachatryan:2016got,Belmont:2016oqp}. 
Those small-system data suggest that the background can be large and, although the physics nature of backgrounds may differ, the $\dg$ in heavy-ion collisions are likely dominated by backgrounds.

An explicit demonstration of ``resonance'' background is the measurement of $\dg$ as a function of the pair invariant mass ($m_{\rm inv}$)~\cite{Adam:2020zsu}. This is shown in Fig.~\ref{fig:minv}, where the $\dg$ is found to trace the \OS\ over \Ss\ pair excess $r\equiv(N_{_{\rm OS}}-N_{_{\rm SS}})/N_{_{\rm OS}}$, large at resonance masses and both on a continuum.

\section{Early Attempts to Address Backgrounds}
There is no question that the background is large.
The question is: How large is the background quantitatively? 
This question has to be answered experimentally when background is large. Many attempts have been made to address the background issue~\cite{Zhao:2019hta,Li:2020dwrReview}.

STAR has measured $\gos$ and $\gss$ in lower energy Au+Au collisions at $\snn=7.7$--62.4~GeV~\cite{Adamczyk:2014mzf}, and the $\dg$ is found to decrease toward lower energies.
Inspired by $\dg_{112}\equiv\dg\approx\mean{\cos(\phia-\phib)}\mean{\cos2(\phib-2\Psi_2)}\approx\kappa_2\Delta\delta v_2$ where $\Delta\delta\equiv\mean{\cos(\phia-\phib)}$~\cite{Bzdak:2012ia}, background estimate using the $\kappa_2$ parameter was attempted. However, the above trigonometry factorization is generally invalid (because of the $\phib$ in both factors)--otherwise the $\kappa_2$ should be unity. The correct factorization is given by Eq.~(\ref{eq:bkgd}). Because the $\kappa_2$ parameter is uncontrolled, rigorous conclusions cannot be drawn on \cme.

It was suggested~\cite{Sirunyan:2017quh,ALICE:2020siw} that $\kappa_3\equiv\dg_{123}/(\Delta\delta v_3)$, where $\dg_{123}$ is the \OS--\Ss\ difference of $\mean{\cos(\phia+2\phib-3\Psi_3)}$, may be a good estimator of the background since no \cme\ exists w.r.t.~the third-order harmonic plane $\Psi_3$. The $\kappa_2$ and $\kappa_3$ were indeed measured to be similar. However, from a $\dg_{123}$ factorization similar to Eq.~(\ref{eq:bkgd}), it is clear that the backgrounds $\kappa_3$ and $\kappa_2$ do not equal. It is thus uncertain how to estimate \cme\ background by $\kappa_3$.

\begin{figure}[htb]
  \begin{minipage}{0.49\textwidth}
    \centering
    \includegraphics[width=0.74\textwidth]{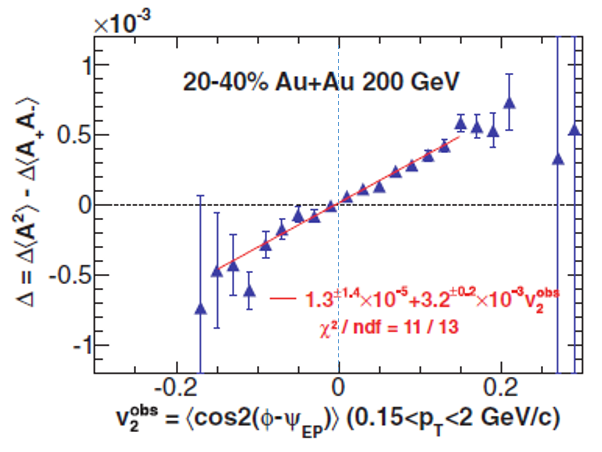}
    \includegraphics[width=0.78\textwidth]{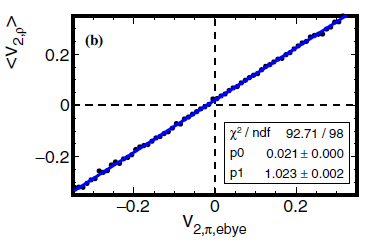}
    \caption{STAR \ese\ analysis~\cite{Adamczyk:2013kcb} engineering on statistical fluctuations of $v_2$ (upper), and toy model study~\cite{Wang:2016iov} of average $v_2$ of the $\rho$ resonance vs.~event-by-event pion $v_2$ in the final state (lower).}
    \label{fig:starese}
  \end{minipage}\hfill
  \begin{minipage}{0.49\textwidth}
    \centering
    \includegraphics[width=0.74\textwidth]{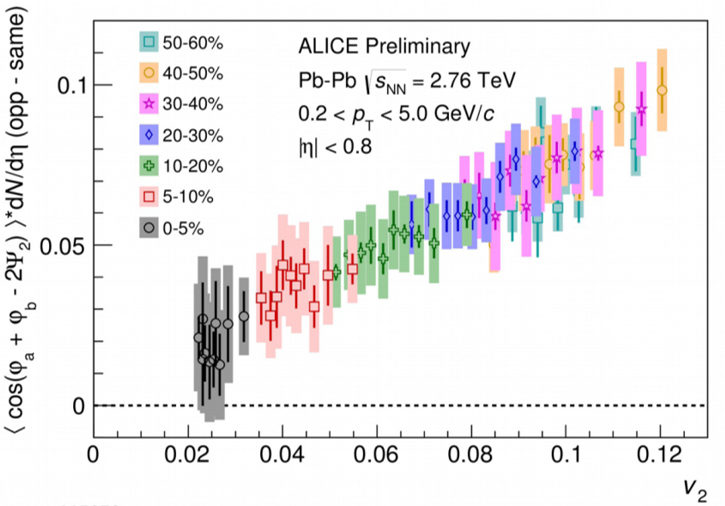}
    \includegraphics[width=0.72\textwidth]{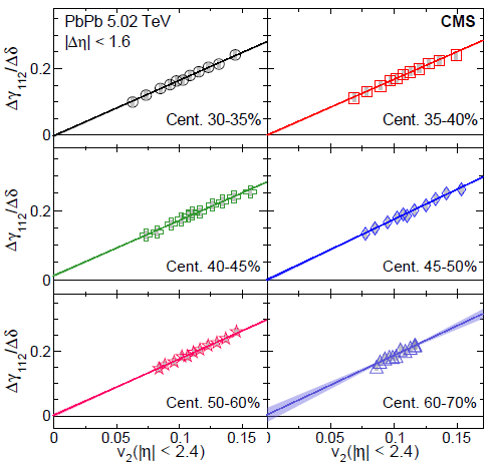}
    \caption{ALICE (upper)~\cite{Acharya:2017fau} and CMS (lower)~\cite{Sirunyan:2017quh} \ese\ analyses engineering on dynamical fluctuations of $v_2$.}
    \label{fig:ese}
  \end{minipage}
\end{figure}

Since $\dg_{\bkg}\propto v_2$, it is attempting to engineer on event shape (\ese) with vanishing $v_2$. This was first attempted by STAR~\cite{Adamczyk:2013kcb} where the $N$-asymmetry correlation (a quantity similar to $\dg$) is plotted against the observed $v_2^{\rm obs}=\mean{\cos(\phi_{_{\rm POI}}-\Psi_{_{\rm EP}})}$ of particles of interest (\poi) in one half of the detector w.r.t.~the \EP\ from the other half. Similar technique has been recently proposed~\cite{Wen:2016zic}. A linear relationship is observed as shown in Fig.~\ref{fig:starese} upper panel; the $v_2^{\rm obs}$ can be negative because this \ese\ method engineers primarily on statistical fluctuations of $v_2$. 
The intercept at $v_2^{\rm obs}=0$, consistent with zero in Fig.~\ref{fig:starese}, would be more sensitive to \cme.
However, it has been found that residual background remains because resonance/cluster $v_2$, primarily responsible for the \cme\ background, does not vanish at $v_2^{\rm obs}=0$ as shown by model study in Fig.~\ref{fig:starese} lower panel~\cite{Wang:2016iov}. 

ALICE~\cite{Acharya:2017fau} and CMS~\cite{Sirunyan:2017quh} have performed the \ese\ analysis by binning events according to the $q_2$ flow vector in the forward/backward region and then studying $\dg$ as a function of the average $v_2$ of \poi\ in those events in each centrality bin~\cite{Schukraft:2012ah}. The results are shown in Fig.~\ref{fig:ese}, where linear relationships are observed. While analysis details differ, both experiments found vanishing intercepts at $v_2=0$, suggesting null \cme\ signals. 
This method engineers on dynamical fluctuations of $v_2$, and remains a promising means to extract the possible \cme, with nonflow effects to be assessed. 

\section{Latest Measurements}
Since background is dominant, in order to extract the possible small \cme\ signal, a delicate ``cancellation'' of background would be required, for which experiments often resort to comparative measures. Two such comparative measures have been carried out recently: one is isobar collisions and the other is correlations w.r.t.~spectator plane (\SP) and participant plane (\PP).

{\em Isobar collisions.}
Isobar collisions were proposed~\cite{Voloshin:2010ut,Skokov:2016yrj} as an ideal means to cancel background: the same mass number of $^{96}_{44}\ru$ and $^{96}_{40}\zr$ would ensure the same background, and the larger atomic number in the former would yield a $\sim$15\% stronger \cme\ signal. This is supported by~model calculations even including nuclear deformations~\cite{Deng:2016knn}. If the \cme\ is 10\% of the measured $\dg$, then an isobar difference of 1.5\% would be expected, representing a $4\sigma$ effect with the precision of 0.4\% achieved in experiment~\cite{STAR:2021mii}. However, because $\dg_{\bkg}\propto1/N$ and the magnetic field is smaller in isobar collisions than in Au+Au, the signal to background ratio in the former may be significantly smaller~\cite{Feng:2021oub}, which would result in a weaker significance. 

Moreover, it has been shown by density functional theory ({\sc dft}) calculations that the isobar nuclear structures are not identical--even though the charge radius of Ru is bigger, Zr possesses a significantly thicker neutron skin leading to its larger overall size~\cite{Xu:2017zcn,Xu:2021vpn}. This would yield larger $N$ and $v_2$ in Ru+Ru than Zr+Zr collisions at the same centrality. As a result, the backgrounds would be slightly different, with an uncertainty that may not be negligible, reducing the significance of isobar collisions~\cite{Xu:2017zcn}. Indeed, the isobar data show significant differences in $N$ and $v_2$ between the two systems~\cite{STAR:2021mii}, 
consistent with {\sc dft} predictions~\cite{Xu:2017zcn,Li:2018oec}.

\begin{figure}[htb]
\centering
\includegraphics[width=0.9\textwidth]{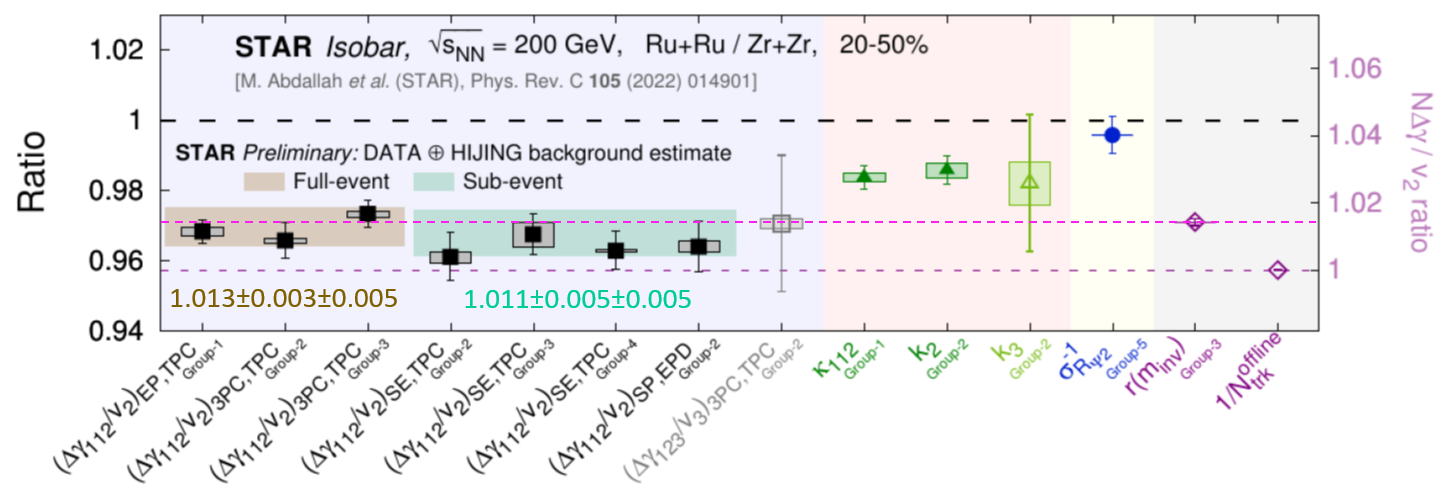}
\caption{The Ru+Ru/Zr+Zr ratios of various \cme\ observables by STAR~\cite{STAR:2021mii,Tribedy:QM}.} 
\label{fig:isobar}
\end{figure}

Figure~\ref{fig:isobar} shows the Ru+Ru/Zr+Zr ratio of various \cme\ observables from STAR~\cite{STAR:2021mii,STAR:2019bjg,Tribedy:QM}. 
The ratio in $\dg/v_2$ being significantly below unity is due to the $N$ difference.
The proper baseline would be the ratio in $1/N$, or unity for the ratio in $N\dg/v_2$, the brown dashed line in Fig.~\ref{fig:isobar}. The $\dg$ data points are all above this line. This, however, cannot lead to the conclusion of a finite \cme\ signal because of nonflow effects~\cite{Feng:2021pgf}. 

The nonflow effects have two parts~\cite{Feng:2021pgf}. 
One is simply because the measured $v_2$ contains nonflow (denoted as $v_2^*$ henceforth, whereas those without asterisk now refer to the true flow), so it is propagated via $N\dg/v_2^*\equiv NC_3/v_2^{*2}$ (similarly in the \EP\ method). 
The other is genuine 3-particle (3p) correlations, the last term in 
$C_3\equiv\mean{\cos(\phia+\phib-2\phic)}=\frac{C_{\twp}N_{\twp}}{N^2}v_{2,\twp}v_2+\frac{C_{\thp}N_{\thp}}{2N^3}$,
where $N\approx N_{+}\approx N_{-}$ is \poi\ multiplicities, $C_{\twp}=\mean{\cos(\phia+\phib-2\phi_{\twp})}$ and $C_{\thp}=\mean{\cos(\phia+\phib-2\phic)}_{\thp}$.
Writing $v_2^{*2}=v_2^2+v_{2,\nf}^2$ (where $v_{2,\nf}^2$ is nonflow contribution), and using shorthand notations
$\enf\equiv v_{2,\nf}^2/v_2^2$,
$\etw\equiv\frac{C_{\twp}N_{\twp}}{N}\cdot\frac{v_{2,\twp}}{v_2}$ and
$\eth\equiv\frac{C_{\thp}N_{\thp}}{2N}$, we have
\begin{equation}
  \frac{(N\frac{\dg}{v_2^*})^{\ru}}{(N\frac{\dg}{v_2^*})^{\zr}}\approx\frac{\etw^{\ru}}{\etw^{\zr}}-\frac{\Delta\enf}{1+\enf}+\frac{\frac{\eth/\etw}{Nv_2^2}}{1+\frac{\eth/\etw}{Nv_2^2}}\left(\frac{\Delta\eth}{\eth}-\frac{\Delta\etw}{\etw}-\frac{\Delta N}{N}-\frac{\Delta v_2^2}{v_2^2}\right)\,.
  \label{eq:isobar}
\end{equation}
Here $\Delta X\equiv X^{\ru}-X^{\zr}$, and variables without superscript refer to those in individual systems $X\approx X^{\ru}\approx X^{\zr}$.
The first term in Eq.~(\ref{eq:isobar}) r.h.s.~characterizes deviation from $N$ scaling--the background scales with $N_{\twp}/N^2$ rather than simply $1/N$. This implies that the baseline should be the ratio of $r$, 
the pink dashed line in Fig.~\ref{fig:isobar}~\cite{STAR:2021mii}.
The nonflow $\enf$ can be assessed by $(\Delta\eta,\Delta\phi)$ 2p correlations. Preliminary data 
indicate a good cancellation between the effect of $v_2$ nonflow in the second term (positive, because $\Delta\enf<0$ due to the larger $N$ in Ru+Ru) and the effect of 3p correlations in the third term (negative); both are of the magnitude 0.5--1\%.
The estimated baselines are indicated by the shaded bands in Fig.~\ref{fig:isobar}.

{\em Measurements w.r.t.~spectator and participant planes.}
Another comparative measure is $\dg$ w.r.t.~\SP\ and \PP\ in the same event~\cite{Xu:2017qfs,Voloshin:2018qsm}. Because the magnetic field is more closely connected to \SP\ and the flow to \PP, these two $\dg$ measurements can uniquely determine the \cme\ and the background. With $\frac{\dgbc{Bkg}{SP}}{\dgbc{Bkg}{PP}}=\frac{\dgbc{_{\rm CME}}{PP}}{\dgbc{_{\rm CME}}{SP}}=\mean{\cos2(\psipp-\psisp)}\equiv a$, one may obtain the \cme\ fraction as 
$\fcme\equiv\frac{\dgbc{_{\rm CME}}{PP}}{\dgbc{}{PP}}=\frac{A/a-1}{1/a^2-1}$ where $A=\frac{\dgbc{}{SP}}{\dgbc{}{PP}}$.

\begin{figure}[htb]
\centering
\includegraphics[width=0.9\textwidth]{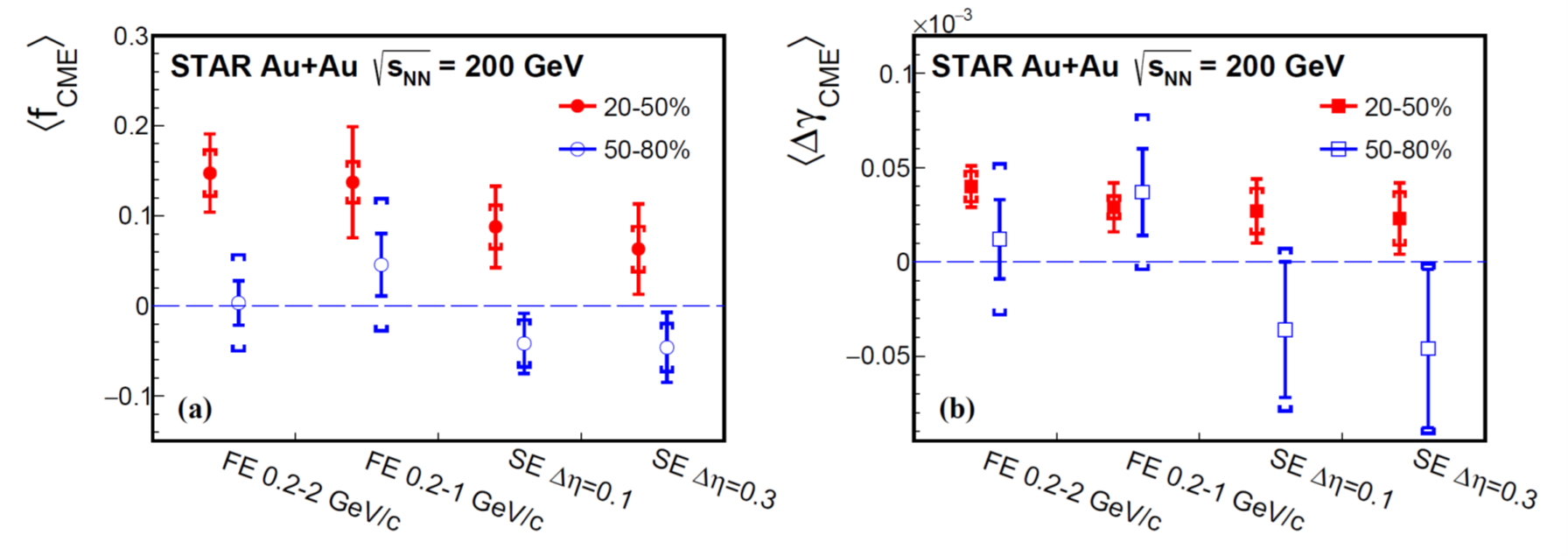}
\caption{The extracted \cme\ fraction $\fcme$ (left) and \cme\ signal $\dg_{_{\rm CME}}$ (right) from $\dg$ measurements w.r.t.~\SP\ and \PP\ in Au+Au collisions at 200~GeV by STAR~\cite{STAR:2021pwb}.}
\label{fig:pprp}
\end{figure}

Figure~\ref{fig:pprp} shows the extracted $\fcme$ and $\dg_{_{\rm CME}}$ in Au+Au collisions at $\snn=200$~GeV by STAR~\cite{STAR:2021pwb}. The peripheral data are consistent with zero \cme\ with relatively large uncertainties. The mid-central 20-50\% data indicate a finite \cme\ signal, with $\sim$$2\sigma$ significance.
Similar analysis has been performed on 27~GeV data, showing zero \cme\ with present statistics~\cite{Tribedy:QM}. 

Similar to isobar collisions, the \SP/\PP\ method measures the ratio of two quantities,
$A/a=\frac{N\dgbc{}{SP}/\vtw{SP}}{N\dgbc{}{PP}/\vtw{PP}}$.
Simpler than the isobar data, only the \PP\ measurements are contaminated by nonflow, $A/a=(1+\enf)/\left(1+\frac{\eth/\etw}{N\vtw{PP}^2}\right)$.
Nonflow in $v_2$ yields a positive $\fcme$ while 3p correlations result in a negative $\fcme$. 
There is a good level of cancellation between the two, and the net effect could even be negative~\cite{Feng:2021pgf}. Although model dependent, it suggests that the measured positive $\fcme$ in data might indeed be a hint of \cme.

\section{Summary and Outlook}
In summary, measurements of the charge correlator $\dg$ and its variants are reviewed. The $\dg$ measurements are dominated by backgrounds arising from genuine particle correlations coupled with elliptic flow $v_2$. Several methods have been devised to eliminate those backgrounds, including event-shape engineering, isobar collisions, and measurements w.r.t.~spectator and participant planes. While the first two yield a \cme\ signal consistent with zero with the present statistics, the third indicates a hint of the possible \cme\ in Au+Au collisions with $\sim$$2\sigma$ significance. All these methods are subject to nonflow effects, the magnitudes of which are under active investigation.

To outlook, an order of magnitude statistics is anticipated of Au+Au collisions from 2023 and 2025 by STAR. This would present a powerful data sample to either identify the \cme\ or put a stringent upper limit on it.
\\
\indent{\bf Acknowledgments.} This work was supported by the U.S.~Department of Energy (No.~DE-SC0012910) and the China National Natural Science Foundation (Nos.~12035006, 12075085, 12147219). 


\end{document}